\magnification=1200
\hoffset=0.1truein
\voffset=0.1truein
\vsize=23.0truecm
\hsize=16.25truecm
\parskip=0.2truecm
\def\newpage{\vfill\eject}

\def\ga{\mathrel{\mathpalette\fun >}}
\def\fun#1#2{\lower3.6pt\vbox{\baselineskip0pt\lineskip.9pt
  \ialign{$\mathsurround=0pt#1\hfil##\hfil$\crcr#2\crcr\sim\crcr}}}
%
%
\rightline{UM-AC-95-11}
\rightline{hep-ph/9512211}
$\,$
\vskip 1.0truein
\centerline{\bf MODULI INFLATION WITH LARGE SCALE STRUCTURE}
\centerline{\bf PRODUCED BY TOPOLOGICAL DEFECTS}
\vskip 0.2truein
\centerline{ {\bf Katherine Freese, Tony Gherghetta, and Hideyuki Umeda} }
\vskip 0.1truein
\centerline{\it Physics Department,
University of Michigan}
\centerline{\it Ann Arbor, MI 48109-1120}
 
\vskip 0.4truein
 
\vskip 0.4truein
 
\centerline{\bf ABSTRACT}
\vskip 0.2truein
\baselineskip=14pt

It is tempting to inflate along one of the many flat directions
that arise in supersymmetric theories. The required flatness
of the potential to obtain sufficient inflation and to not 
overproduce density fluctuations occurs naturally.
However, the density perturbations (in the case of a single
moduli field) that arise from inflaton quantum fluctuations are 
too small for structure formation.
Here we propose that topological defects (such as cosmic strings), 
which arise during a phase transition near the end of moduli inflation 
can provide an alternative source of structure.  The strings produced 
will be `fat', yet have the usual evolution by the time of 
nucleosynthesis. Possible models are discussed. 
 
\newpage
 
{\bf I. Introduction}

The inflationary universe model was proposed [1]
to solve several cosmological puzzles, namely 
the horizon, flatness, and monopole
problems. During the inflationary epoch,
the energy density of the universe is dominated by the
vacuum energy, $\rho \simeq \rho_{vac}$,
and the scale factor of the universe expands superluminally.
In many models this expansion is exponential,
$R(t) \propto e^{Ht}$, where the Hubble
parameter $H$ = $\dot R /R
\simeq (8 \pi \rho_{vac}/(3 m_{pl}^2))^{1/2}$ during inflation 
and $m_{pl}\sim 10^{19}$GeV is the Planck mass.
If the interval of exponential expansion satisfies
$\Delta t \ga 60 H^{-1}$, a small causally
connected region of the universe grows sufficiently large 
to explain the observed homogeneity and isotropy
of the universe. In addition, any overdensity 
of magnetic monopoles is diluted.
 
To satisfy a combination of constraints on inflationary models [2],
in particular, sufficient inflation and microwave background anisotropy 
limits [3] on density fluctuations, the potential of the field responsible 
for inflation (the {\it inflaton}) must be very flat.
It was shown in [7] that, 
for a general class of inflation
models involving a slowly-rolling field
(including new [4], chaotic [5], and double field [6] inflation),
any potential satisfying these two constraints together with the
condition of overdamping must also obey the following condition
$$\chi \equiv \Delta V/(\Delta \psi)^4
\le {\cal O}(10^{-6} - 10^{-8}) \, .
\eqno(1)$$
Here $\chi$ is the ratio of the height to the (${\rm width})^4$ 
of the potential, i.e.,
$\Delta V$ is the change in the potential $V(\psi)$
and $\Delta \psi$ is the change in the inflaton field $\psi$
during the slowly rolling portion of the inflationary epoch. 
Thus, the couplings in the inflationary potential must be small;
for example, if the inflationary potential is quartic,
then the quartic coupling constant must satisfy 
$\lambda < {\cal O}(\chi)$.

Introducing arbitrary small couplings at tree level in the inflationary 
potential is unnatural because a
fine-tuning must be performed to cancel large radiative corrections.
This procedure would simply replace a cosmological naturalness problem 
with unnatural particle physics. Instead, there are two different 
attitudes one can take to explain this required small number.  
One can simply resort 
to future physics: we know that there is a hierarchy problem (e.g., 
the mass of the electroweak
Higgs is much smaller than the grand unified scale), and hopefully
expect that whatever solves the hierarchy problem might someday explain the
small ratio of scales required for inflation.  Alternatively,
one can look for small numbers in particle physics today.  One 
possibility, that has been
explored in the Natural Inflation model [8], is to identify the inflaton 
as a Nambu-Goldstone boson. Another possibility
is to take advantage of supersymmetry and invoke the 
`technical naturalness' argument, where small numbers once postulated 
at tree level in the superpotential, are protected by supersymmetry 
from receiving large radiative corrections [9].  

Recently, there has been interest in trying to inflate along (nearly) flat 
directions in supersymmetric field theories [10-12]. Flat directions are 
directions in field space where the classical scalar potential exactly 
vanishes and are parametrised
by complex scalar fields, referred to as moduli fields, $\psi$.
In the supersymmetric limit the potential along these flat directions
vanishes identically (neglecting nonrenormalisable terms), i.e., 
$V(\psi) = 0$. However soft supersymmetry breaking terms will lift the 
scalar potential by an amount 
$V(\psi) = m_W^2 |\psi|^2$, where $m_W$ must be of order the electroweak
scale to solve the hierarchy problem associated with the electroweak Higgs 
mass (all numerical values in the paper are obtained with $m_W\sim 1$ TeV). 
In the inflationary context this potential is still 
very flat because $m_W \ll m_{pl}$, where typically $\Delta\psi \sim 
{\cal O}(m_{pl})$ in the early universe.
Thus the constraint in Eq. (1) on the ratio $\chi$ is easily satisfied.

We will consider an inflationary epoch where the inflaton is identified 
with a moduli field, $\psi$, and the inflationary potential is given by the
soft-supersymmetry breaking term $V(\psi) = m_W^2 |\psi|^2$. The moduli field
has an initial value $\psi_0\sim 4-5 \, m_{pl}$ (as in chaotic inflation [5])
and the universe inflates 
as the field $\psi$ rolls down the potential. 
Moduli inflation using soft terms was previously 
discussed in Refs [11,12]. 
An interesting consequence of moduli inflation, pointed out by Randall
and Thomas [12], is that one can avoid the `cosmological moduli' problem [13].
Normally, weakly coupled scalar fields with masses $m \ll H$ 
and initial values of ${\cal O}(m_{pl})$ that are displaced far from
their minima either overclose the universe, or decay so late that they 
destroy the predictions of nucleosynthesis. 
This problem is resolved by a period of moduli inflation because
typical scalar masses
$m\sim m_W\sim H$ and the offending scalar fields 
are quickly driven to their minima. Possible caveats to this solution 
have been addressed by [14]
(e.g., there may still be a residual moduli problem if the potential
minima do not coincide before and after inflation), but scenarios exist where 
this approach could work.

However, a problem that arises during inflation with a single
moduli field is that the magnitude of the density perturbations
produced is {\it too small}. This can easily be seen by 
considering the equation of motion for the scalar field during inflation,
$$\ddot{ |\psi|} + 3 H \dot{|\psi|} = - {dV \over d|\psi|} \, . \eqno(2)$$
In the overdamped approximation known as `slowly rolling' one
may neglect the acceleration term $(\ddot{ |\psi|})$ during inflation. 
In general the density fluctuations scale with the height of the
potential and for a model of inflation
driven by the potential $V = m_W^2 |\psi|^2$, we obtain
$${\delta \rho \over \rho} \sim {1 \over 10} {H^2 \over \dot{ |\psi|}}
\sim {m_W \over m_{pl}^3} |\psi|^2 \, . \eqno(3)$$  
In the early universe, a typical value for the scalar field
is $\psi \sim m_{pl}$, and so the density fluctuations produced are roughly
$${\delta \rho \over \rho} \sim {m_W \over m_{pl}} \sim 10^{-16} \, . 
\eqno(4)$$
This value is too small to explain the observed large scale
structure.  Recent COBE measurements of
microwave background anisotropies obtain a value [3]
$${\delta \rho \over \rho}|_{obs} = {\rm few} \times 10^{-5} \, . \eqno(5)$$

This general problem of producing
large enough density perturbations for moduli inflation occurs because
the known scales in particle physics do not coincide with 
the scale needed for density perturbations.
Infact, in a recent moduli
inflation model by Thomas [10], a dynamical supersymmetry breaking 
scale is introduced at $\Lambda\sim 10^{16}$GeV solely for the purpose
of producing the correct density perturbations. Unfortunately, 
supersymmetry breaking at $10^{16}$GeV has no relevance for the 
physical particle spectrum and supersymmetry needs to be
restored at the end of inflation.
If we do use relevant soft terms for the inflationary potential,
then the 
density fluctuations are too small. This is because the height of the 
potential is too small. In Ref. [12], the lack of sizeable 
density perturbations is avoided by assuming that moduli inflation 
is preceded by an earlier inflationary epoch that produces the correct
magnitude of density perturbations. In order not to wipe out these
density perturbations the subsequent moduli inflationary period can 
only last for at most 30 e-folds. In recent work Randall and Guth [21]
have been working on coupling two scalar fields (with a potential we 
describe in section IIC) to obtain a hybrid inflation model [26,27] with 
adequate density fluctuations.

Here we propose, instead, that the density fluctuations
responsible for the formation of large scale structure are produced 
from cosmic topological defects such as cosmic strings [15].  
Near the end of 
the $\psi$ moduli field driven inflation (or after
inflation), a phase transition is
induced in another complex scalar field $\phi$, which 
creates cosmic defects. The term in the Lagrangian
that drives the phase transition is of the form $H^2 |\phi|^2$;
such a term is necessarily always present in the early universe.
Cosmic strings arise when a U(1) symmetry is spontaneously broken, 
which occurs when the mass squared $(m_\phi^2)$ term
changes sign.
In general, the radius of the string core is given by [16]
$$\delta_\phi \sim m_\phi^{-1} \, . \eqno(6)$$
As we will show, the cosmic strings produced in this model
are fatter than usual by a factor of $10^{10} - 10^{13}$.
However, it turns out that by the time the strings play
any role in physics that might be observable, such as
during nucleosynthesis or at recombination, the universe
is sufficiently large that the thickness of the strings is
again negligible.  The size of the fat strings is roughly
$10^{-20} - 10^{-17}$ cm, while the horizon size at nucleosynthesis
is $\sim 10^{10}$ cm.  Thus, the strings behave as usual for any
observables (and for the formation of cosmic structure).

The only parameter that enters into the formation of cosmic
structure is the mass per unit length of the string $\mu$,
which must have a value $ \mu \sim 10^{-6} m_{pl}^2$.
There are two types of cosmic strings possible depending on 
whether the U(1) symmetry is local or global. 
For local strings (e.g., for a cosmic string potential of the form 
$V_\phi = \lambda (\phi^\dagger\phi - \eta^2)^2$)  the 
string has an inner core with linear mass density [16]
$$\mu \sim \eta^2 \, , \eqno(7)$$
where $\eta$ is the minimum of the cosmic string potential.
Requiring $ \mu \sim 10^{-6} m_{pl}^2$ then determines
the minimum of the string field potential to be $\eta \sim 10^{16}$GeV.  
In the case of global strings, one obtains instead 
$$\mu \sim 2 \pi \eta^2 {\rm ln}(R/\delta_\phi) \, , \eqno(8)$$
where $R$ is a cutoff given either by the radius of the
string loop or by the distance to the neighboring string.
For global strings parametrised by (8), the location of the minimum is 
roughly $\eta \sim 10^{15}$GeV for fat strings.  
Hereafter for simplicity, we will only consider examples of string
potentials with a global U(1) symmetry.  We will impose the
condition that after the phase transition the string field sits
at a minimum $\langle \phi \rangle \sim 10^{15}$ GeV, so that global
cosmic strings can explain the observed density fluctuations.

We should comment that many authors have been working on a
comparison of predictions from cosmic
strings and textures with various observations, including
the microwave background and the power spectrum for large scale
structure.  For example, Crittenden and Turok [25] have
pointed out that textures will produce a Doppler peak in 
the microwave background at scale $l \sim 400$
(whereas inflation should produce a peak at $l \sim 200$.)
Whether or not cosmic defects will prove to be in concordance
with upcoming observations and will consequently provide the explanation
for the origin of large scale structure is of course at present unclear.

Note that the idea of cosmic string production during or near an inflationary
era is not new and has been considered by a number of authors [15]. Early work
on this subject includes a paper by Shafi and Vilenkin [15] who showed that 
the spontaneous breaking of a global U(1) symmetry in minimal SU(5) grand 
unification can produce topologically stable strings at the end of an 
inflationary era. Various scenarios for coupling the string field to 
the inflaton such as via a direct coupling of the two fields or via the 
spacetime curvature scalar have also been considered [15]. However, in this 
previous work the formation of topological defects was considered in the 
context of inflation with a Hubble constant $H >> m_W$. In the present work 
we are considering topological defects in the interesting context of moduli 
inflation where $H\sim m_W$.

The plan for the rest of the paper is as follows:
In Section II we consider a model of moduli
inflation in which the large scale structure is formed by cosmic defects.
We then discuss the various constraints any such model must satisfy, 
and illustrate the resulting requirements for parameters in the
model. We will present three different examples of the cosmic string 
potential and comment on the better motivated scenarios.
Further discussion and our conclusion will be given in Section III.

{\bf II. Models of moduli inflation with cosmic strings}

Consider two complex scalar fields $\psi$ and $\phi$. We assume that
the field responsible for inflation is a moduli field, $\psi$, which
has a soft supersymmetry-breaking potential.
The second field, $\phi$, undergoes a phase transition near
the end of inflation and gives rise to cosmic defects;
for definiteness, we will take cosmic strings as an example.
The potential for these two fields is assumed to have the form
$$V = m_W^2 |\psi|^2 + c H^2 |\phi|^2 + V_\phi \, . \eqno(9)$$
The last term, $V_{\phi}$ is 
the potential for the cosmic string field and is responsible 
for producing the symmetry breaking minima.
The second term is always present in the early universe for
any scalar field and arises from considering the full scalar potential
of N=1 supergravity. This contribution to the $\phi$ scalar field mass
may in general be of either sign.  For example, as discussed by [17]
a negative contribution will arise from the Kahler potential term
$\delta K \sim (1/m_{pl}^2) \psi^\dagger \psi\phi^\dagger\phi$.
However, we will assume that the value of the coupling $c$ is positive 
and of order one ($c$=3 for a minimal Kahler potential). Note that a 
similar term, $H^2\psi^2$ arises for the inflaton field [27,28], but since 
$H\sim m_W$ for moduli inflation as noted in the introduction, this term is 
comparable to the soft-breaking terms already present in Eq.(9).

Note that one could consider an additional interaction term
in the Lagrangian $g^2 |\phi|^2 |\psi|^2$, which would contribute
an effective mass term for $\phi$. This would typically 
dominate over the $H^2 |\phi|^2 $ term, and become responsible for 
the symmetry breaking of the string field. The details of the string 
production in this scenario depend on the values of the parameters and 
requires a more thorough investigation. It is also possible that 
if $g\sim {\cal O}(1)$, thermal effects during the reheating stage of the 
universe generate $T^2 \phi^2$ terms which will trap the string field at 
the origin leading to a thermal inflation phase [18].
In this case the cosmic string production occurs after the universe cools 
to a temperature $T\sim m_\phi$. Cosmic strings could then form quite 
late, e.g., at the electroweak scale. However this thermal effect can be 
avoided if the coupling, $g$ is too weak to allow thermalisation. For the 
remainder of this paper we do not consider this interaction term further.

The basic evolution of both fields is as follows:
The inflaton field $\psi$ starts out
at a value $\geq m_{pl}$ and is assumed to dominate the energy density of 
universe. An inflationary epoch commences as $\psi$ rolls down towards 
its minimum at the origin. Since the vacuum energy during inflation
$\rho_V \sim m_W^2 m_{pl}^2$, the Hubble constant $H =
[8 \pi \rho_V/ (3 m_{pl}^2)]^{1/2}
\sim m_W$.
The mass of the string field $\phi$ is assumed to be dominated by
the contribution $H^2 |\phi|^2$. During inflation this field will
be quickly driven towards the origin.
As inflation proceeds, $H$ will slowly decrease and
at some point near the end of inflation,
negative mass squared terms in $V_\phi$ will begin to dominate.
This causes a phase transition and $\phi$ falls towards
its new minimum (assumed to be at $m_{GUT}$).  
Cosmic strings (or other defects)
are created in the process and will then become responsible for the 
formation of structure.  Note that the density
fluctuations produced directly from the inflaton quantum fluctuations 
are too small to play any role.  The solution to the cosmological
moduli problem as well as reheating proceed in the same
way as discussed in [12].

Now we present three different possible potential terms, $V_\phi$
for the string field and discuss the constraints on each
possibility.

{\bf IIA}. 
Consider first the scalar potential
$$V_{\phi} = \lambda (\phi^\dagger\phi - \eta^2)^2 \, , \eqno(10)$$
which is similar to a supersymmetric GUT Higgs potential.
The radius of the resultant strings follows from Eq. (6)
and is given by
$\delta_\phi \sim m_\phi^{-1} \sim \lambda^{-1/2} \eta^{-1}$;
the mass per unit length is
$\mu \sim \eta^2$.
As mentioned in the Introduction, requiring $\mu \sim 10^{-6} m_{pl}^2$
determines $\eta \sim 10^{15}$ GeV.  
The constraint on the model are as follows.

\noindent
1.  The energy density must be dominated by the inflaton field $\psi$.
Thus the vacuum energy density of the string field must satisfy
$$\lambda \eta^4 <  m_W^2 |\psi|^2 \,  \eqno(11)$$
during the inflationary epoch.
Since $\psi \sim m_{pl}$, this means that
$\lambda < 10^{-16}$. Although such a small number may be
`technically natural', the potential
(10) with an extremely small $\lambda$ lacks motivation.

\noindent
2.  There must be symmetry breaking of the U(1) associated
with the string field $\phi$ in order to generate the strings.  
This happens when the mass squared term of the $\phi$ field changes
sign, i.e., when $ \lambda \eta^2 \sim H^2$. 
Since $H\sim (m_W/m_{pl})|\psi| $ decreases during
inflation, this criterion can be eventually reached.
Strings can be produced any time after 50 e-folds 
before the end of inflation [15]; then the strings are not diluted 
too much by the subsequent inflation to be of relevance for
structure formation. [Note that the $\lambda\eta^4$ term does not 
affect when the phase transition occurs.]
Thus, the coupling must satisfy
$\lambda \leq 10^{-24},$ 
where the upper bound corresponds to cosmic strings forming
near the end of inflation.
This value is even smaller than that required by the first constraint,
and as discussed above, such a small number is unmotivated.

In the next two examples
we study two scenarios with potentials qualitatively 
similar to Eq. (10) but not requiring extremely small parameters.

{\bf IIB.}
Here we follow Lyth and Stewart [18] 
and consider
$$V_{\phi} = V_0 - m_W^2 |\phi|^2 +
b {|\phi|^{n+4} \over m_{pl}^n} \, , \eqno(12)$$
where $n>1$ and $b$ is a constant.  
The minimum of the potential (12) occurs at
$$\langle \phi \rangle = \bigg[{2 m_W^2 m_{pl}^n \over
(n+4) b} \bigg]^{1/(n+2)}  \, . \eqno(13)$$
In order to obtain $\langle \phi \rangle 
\sim 10^{15}$ GeV, as required for cosmic string formation with
the correct mass per unit length, we 
need $n \approx 6$, assuming $b$ to be of order one. This 
requires all nonrenormalizable 
terms with $n<6$ to be suppressed; otherwise the minimum will be too low
in energy. This could be possible if one identifies $\phi$ with a
flat direction which is lifted by a dimension 4 superpotential term [19].
Alternatively the situation may not be quite as extreme
if there is a reason to obtain $b \ll 1$.  Then $n$ need
not be as large.  This may, for example, happen in string theory
if one imposes discrete symmetries which only
allow specific couplings of the last term in Eq. (12) 
with remnant string fields, $S$ [20] (note that $S$ does {\it not}
refer to cosmic string fields).  For example,  one may have $b \sim \langle
S\rangle^p/m_{pl}^p$ where $p$ is some integer
and $\langle S \rangle / m_{pl} \sim 0.1$ at the string scale.
In this way one hopes to get a minimum for the potential
at the GUT scale.

The constant term $V_0$ must be added to obtain the right value of the
cosmological constant today, $\Lambda \sim 0$. Requiring $V_\phi=0$
at the potential minimum $\langle \phi \rangle \sim 10^{15}$ GeV
gives $V_0 \sim (10^9 {\rm GeV})^4$ (for $n=6$).
The mass per unit length of the cosmic strings produced
will then be $\mu \sim 10^{-6} m_{pl}^2$
as required.  The thickness of the strings is $\delta_\phi \sim
m_W^{-1} \sim 10^{13} m_{GUT}^{-1}$ where $m_{GUT} \sim
10^{16}$ GeV, i.e., $10^{13}$ times as large as usual. Indeed these 
are fat strings.

The required constraints for inflation followed by cosmic string
production to work can be satisfied.
Indeed the constraint that the inflaton potential dominate
the energy density of the universe is satisfied:
the vacuum energy of the string field $V_0$ is smaller than that of the
inflaton, i.e., $V_0 < m_W^2 m_{pl}^2 \simeq (10^{11}$ GeV)$^4$.
The phase transition in the string field occurs
when $H^2 \sim m_W^2$, i.e., when $\psi \sim m_{pl}$.
Thus, one can have moduli inflation with cosmic string production
near the end of inflation, where both $\psi$ and 
$\phi$ can be identified with flat directions in a 
supersymmetric theory.

{\bf IIC}.
The third possibility we consider is
$$V_\phi = M^4 {\rm cos}^2{|\phi| \over f} \, , \eqno(14)$$
where $M$ is some as yet unspecified mass scale and the
minimum of the potential must be at $f \sim 10^{15}$ GeV in order
to obtain the correct $\mu$ for the cosmic strings.  Unfortunately
such a value for $f$ is not well-motivated.  The same
form of the potential is considered by Randall and Guth [21]
in constructing a hybrid inflation model with moduli
(they do not require the same value of $f$, however).
For $|\phi| \ll f$, we can expand the string potential so that
$$V_\phi = M^4 - {M^4 \over f^2} |\phi|^2 + {1 \over 3} 
{M^4 \over f^4} |\phi|^4 \, .  \eqno(15)$$
Then the constraints on the model are as follows:

\noindent
1. The inflaton field $\psi$ must dominate the energy density.
This means that $M^4 < m_W^2 |\psi|^2$. So,
for $\psi \sim m_{pl}$ during inflation, we need $M \leq 10^{11}$GeV.

\noindent
2. Strings can form when $H^2 \sim {M^4 \over f^2} $, where
$H \sim (m_W/m_{pl}) |\psi|$ during inflation.
Since $f \sim 10^{15}$ GeV is fixed whereas $\psi$ continually
decreases we obtain $M \leq 10^9$ GeV. Such intermediate mass
scales responsible for dynamical supersymmetry  breaking 
are possible.

The string parameters are similar to the previous cases.
The mass per unit length of the cosmic strings is given by
$\mu \sim (10^{16}$ GeV)$^2$,
as required. The thickness of the cosmic strings is 
$\delta_\phi \sim m_\phi^{-1} $ $\sim f/M^2 \sim 10^{13} m_{GUT}^{-1}$,
i.e., $10^{13}$ times as large as the usual strings.
Note that the coefficient in front of the $\phi^4$ term
is ${1 \over 3} {M^4\over f^4} \sim 10^{-25}$,
approximately the same
value that was required in the example studied in Section IIA.

{\bf III. Discussion and Conclusion}

Inflation using soft terms with a single moduli field by 
itself is unsatisfactory
because inadequate density fluctuations are produced.
We have proposed that cosmic defects may be formed at the end of 
an inflationary epoch and provide the large scale structure.  
We have focused on
cosmic strings as an example.  The cosmic
strings that can be produced during moduli inflation 
are `fat' compared to usual strings, 
with thickness ranging from
$10^{10}$ to $10^{13}$ times the  usual values. 
Thus the thickness ranges from $(10^{10} - 
10^{13}) m_{GUT}^{-1} \sim (10^{-20} - 10^{-17})$cm.  However,
the earliest observable effects from the strings would be
produced at nucleosynthesis, by which time even these
fat strings would be `thin' relative to the horizon size,
roughly $10^{10}$cm. [At that time the production of
gravitational waves by the strings might serve to constrain
them very weakly].
Certainly the most likely observable effects would be 
produced subsequent to the time of recombination at $T \sim $ eV,
by which time the initial fatness of the strings would
be completely irrelevant.  The horizon size at
recombination is roughly $10^{20}$ cm.  
Thus these strings follow the usual evolution [22].

If the potential for the string field is minimized at 
$\sim 10^{16}$GeV, then the required value of mass per unit
length of the cosmic strings is obtained.
We examined three different string field potentials:
1) $V_\phi = \lambda (\phi^\dagger\phi - \eta^2 )^2$
required $\lambda \sim 10^{-24}$, which is not very well motivated;
2) $V_\phi = V_0 - m_W^2 |\phi|^2 + b{|\phi|^{n+4}
\over m_{pl}^n}$ needed $n\approx 6$
for $b \sim 1$.  Smaller values of $b$ may be obtained 
from string theory by imposing discrete symmetries, which would
allow more reasonable values of $n$;
3) $V_\phi = M^4 {\rm cos}^2{|\phi| \over f}$ required
$f \sim 10^{15}$ GeV, not a well-motivated value.
While none of these potentials is perfect, we hope that the 
examples presented are illustrative.

We would also like to point out that there are other
ways to produce cosmic strings during inflation.  First, Basu,
Guth and Vilenkin [23] have studied the production of cosmic
defects that arise out of fluctuations of the vacuum during
inflation.  Second, Kofman, Linde, and Starobinsky [24] have
proposed that cosmic defects may be able to arise due to parametric
resonance giving rise to temperature effects that induce
a phase transition during reheating after inflation.
If either of these two mechanisms is active, these
would be alternative ways to generate cosmic defects,
and thereafter large scale structure, in a model
of single moduli inflation.

{\bf Acknowledgements:}
We would like to thank Andy Albrecht, Ken Intriligator, Albert Stebbins,
Ewan Stewart, and Alex Vilenkin for useful discussions 
and/or correspondence. This work would not have seen the light
of day were it not for discussions that took place at the Aspen
Center for Physics, whose hospitality we gratefully acknowledge.
K.F. thanks NSF grant PHY-9407194, T.G. thanks the U.S. Department
of Energy, and H.U. thanks the Physics Department 
at the University of Michigan for support.

\newpage

\vskip 0.80truein
\centerline{\bf REFERENCES}
\vskip 0.20truein
 
\item{[1]} A. H. Guth, {\it Phys. Rev.} {\bf D23}, 347 (1981).
 
\item{[2]} P. J. Steinhardt and M. S. Turner, {\it Phys. Rev.}
{\bf D29}, 2162 (1984).
 
\item{[3]} G.F. Smoot {\it et al}, {\it Astrophys. J. Lett.} 
{\bf 396}, L1 (1992).
 
\item{[4]} A. D. Linde, {\it Phys. Lett.} {\bf B108}, 389 (1982);
A. Albrecht and P. J. Steinhardt, {\it Phys. Rev. Lett.}
{\bf 48}, 1220 (1982).
 
\item{[5]} A. D. Linde, {\it Phys. Lett.} {\bf B129}, 177 (1983).
 
\item{[6]} F. C. Adams and K. Freese, {\it Phys. Rev.} 
{\bf D51}, 6722 (1995).
 
\item{[7]} F. C. Adams, K. Freese, and A. H. Guth,
{\it Phys. Rev.} {\bf D43}, 965 (1991). 
 
\item{[8]} K. Freese, J.S. Frieman, and A.V. Olinto,
{\it Phys. Rev. Lett.} {\bf 65}, 3233 (1990);
F.C. Adams, J.R. Bond, K. Freese, J.A. Frieman, and A.V. Olinto,
{\it Phys. Rev.} {\bf D47}, 426 (1993).

\item{[9]} See, e.g., the review of K. Olive, {\it Phys. Rep.}
{\bf 190}, 307 (1990);
R. Holman, P. Ramond, and G.G. Ross, {\it Phys. Lett.} {\bf B137},
343 (1984);
H. Murayama, H. Suzuki, and T. Yanagida, {\it Phys. Rev. Lett.}
{\bf 70}, 1912 (1993);
T. Gherghetta and G.L Kane, {\it Phys. Lett.} {\bf B354}, 300 (1995).

\item{[10]} T. Banks, M. Berkooz, S.H. Shenker, G. Moore,
and P.J. Steinhardt, {\it Phys. Rev.} {\bf D52}, 3548 (1995);
S. Thomas, {\it Phys. Lett.} {\bf B351}, 424 (1995).

\item{[11]} F. Graziani and K. Olive, {\it Phys. Lett.} 
{\bf B216}, 31 (1989).

\item{[12]} L. Randall and S. Thomas, {\it Nucl. Phys.} 
{\bf B449}, 229 (1995).

\item{[13]} G.D. Coughlan {\it et al}, {\it Phys. Lett.} {\bf B131},
59 (1983); B. de Carlos, J.A. Casas, F. Quevedo, and E. Roulet,
{\it Phys. Lett.} {\bf B318}, 447 (1993); T. Banks, D.B. Kaplan,
and A.E. Nelson, {\it Phys. Rev.} {\bf D49}, 779 (1994).

\item{[14]} T. Banks, M. Berkooz and P.J. Steinhardt, 
{\it Phys. Rev.} {\bf D52}, 705 (1995).

\item{[15]} Q. Shafi and A. Vilenkin, {\it Phys. Rev.} {\bf D29}, 1870 
(1984). L.A. Kofman and A.D. Linde, {\it Nucl. Phys.} {\bf B282}, 555 (1987);
E. Vishniac, K.A. Olive, and D. Seckel, {\it Nucl. Phys.}
{\bf B289}, 717 (1987);  D. H. Lyth {\it Phys. Lett.} {\bf B246}, 359
(1990); J. Yokoyama, {\it Phys. Rev. Lett.} {\bf 63}, 712 (1989);
H. Hodges and J. Primack, {\it Phys. Rev.} {\bf D43}, 3161 (1991); 
M. Nagasawa and J. Yokoyama, {\it Nucl. Phys.} {\bf B370}, 472 (1992).

\item{[16]} A. Vilenkin, {\it Phys. Rep.} {\bf 121}, 263 (1985).

\item{[17]} M. Dine, L. Randall, and S. Thomas,
{\it Phys. Rev. Lett.} {\bf 75}, 398 (1995); 
G. Dvali, {\it Phys. Lett.} {\bf B355}, 78 (1995).

\item{[18]} D.H. Lyth and E.D. Stewart, 
{\it Phys. Rev. Lett.} {\bf 75}, 201 (1995);
{\it Phys. Rev.} {\bf D53}, 1784 (1996).

\item{[19]} T. Gherghetta, C. Kolda and S.P. Martin, hep-ph/9510370,
{\it Nucl. Phys.} {\bf B468}, 37 (1996).

\item{[20]} G. Dvali and Q. Shafi, {\it Phys. Lett.} {\bf B339},
241 (1994).

\item{[21]} L. Randall and A. Guth, private communication.

\item{[22]} We thank Albert Stebbins for discussions on the evolution
of fat strings. Fat strings have also been discussed by G. Lazarides,
C. Panagiotakopoulos and Q. Shafi, {\it Phys. Rev. Lett.} {\bf 56}, 432 
(1986).

\item{[23]} R. Basu, A.H. Guth, and A. Vilenkin, {\it Phys. Rev.}
{\bf D44}, 340 (1991).

\item{[24]} L. Kofman, A. Linde, and A. Starobinsky, {\it Phys. Rev. Lett.} 
{\bf 76}, 1011 (1996).

\item{[25]} R. Crittenden and N. Turok, {\it Phys. Rev. Lett.} 
{\bf 75}, 2642 (1995).

\item{[26]} A. Linde {\it Phys. Lett.} {\bf B259}, 38 (1991).

\item{[27]} E.J. Copeland, A.R. Liddle, D.H. Lyth, E.D. Stewart and D. Wands,
{\it Phys. Rev.} {\bf D49}, 6410 (1994).

\item{[28]} G. Dvali, Q. Shafi and R. Schaefer, {\it Phys.Rev.Lett.} 
{\bf 73}, 1886 (1994).

\bye